\documentclass[a4paper,12pt]{article}

\usepackage{amsmath, amsfonts, amssymb}
\usepackage{graphicx}
\usepackage[left=2.2cm, right=2.2cm, top=3cm, bottom=3cm]{geometry}

\usepackage[utf8]{inputenc}
\usepackage[T1]{fontenc}
\usepackage[english]{babel}
\usepackage{enumitem}
\usepackage{multirow}

\usepackage{mathtools}

\usepackage[numbers,sort&compress]{natbib}

\usepackage[colorlinks,linkcolor=blue,linktoc=section]{hyperref}

\newtheorem{theorem}{Theorem}
\newtheorem{corollaire}{Corollary}
\newtheorem{lemma}{Lemma}

\newenvironment{proof}{\par\noindent\textbf{Proof.}}{\hfill $\blacksquare$\par\smallskip}

\begin{document}

\begin{sloppypar}

\begin{center}
\textbf{\Large{Integrability of the Schrödinger Equation for a Mie-type Potential via the NU Method and Appell-type Hypergeometric Polynomials}}
\end{center}

\begin{center}
\textrm{\normalsize{\textbf{$^{\ast\S} $Kikunga\  Kasenda\ Ivan, $^\ast$Omalohoso Tasa Joshua ,$^\dagger$Likwolo\  Katamba\ Stanislas}}}\\
\end{center}

\begin{center}
{\textit{\scriptsize{$^\ast$Physics and technology Mention, Faculty of Sciences and Technologies, University of Kinshasa,P.O.Box Kin\ 190,Kinshasa, D.R.Congo.}}}\\
{\textit{\scriptsize{$ ^\S $Department of Engineering Sciences, Faculty of Sciences and Technologies, Université Loyola du Congo, P.O.
Box 3724 Kin I, Kinshasa, D.R.Congo.}}}\\
{\textit{\scriptsize{$^\dagger$Faculty of Agronomy,University of Uele,P.O.Box Isiro\ 670, Isiro, D.R.Congo.}}}\\
\end{center}

\begin{center}
{\scriptsize{Email : ivan.kikunga@unikin.ac.cd, josue.omalohoso@gmail.com, likwolostanislas@yahoo.fr}}\\
\end{center}
\


\begin{abstract}
	\
	
	{\footnotesize{We establish a formal correspondance between the Appell-type hypergeometric polynomials $\widetilde{e}_n$, recently introduced in \cite{Art05-37}, and the exact solutions of the radial Schrödinger equation for a family of Mie-type potentials. Using the Nikiforov-Uvarov (NU) method, we show that the radial equation reduces to a hypergeometric differential equation with a parameter $N$, which plays the role of an effective principal quantum number. A careful analysis of the asymptotic behavior of the solutions reveals a dual structure: the confluent hypergeometric function of the second kind $U$ yields square-integrable bound states, for which the condition $N\in\mathbb{N}$ provides a quantization condition of Diophantine type, and a discrete energy spectrum. In contrast, the generalized Laguerre function $L$, of a certain type, yields non-normalizable scattering states. In the particular case of vanishing angular momentum $(l=0)$, these scattering states truncate into the Appell-type polynomial $\widetilde{e}_n$, which thus characterize the $s-$wave scattering of the Mie-type potential. This work therefore provides a unified algebraic framework for both bound and scattering states, and clarifies the physical scope of the Appell-type polynomials.}}
\end{abstract}
\


\section{Introduction}
\label{sec01}
\

The resolution of the Schrödinger equation for exactly solvable potentials remains a central problem in non-relativistic quantum mechanics. Among the various techniques available, the Nikiforov-Uvarov (NU) method\cite{Art05-10} stands out as a unified framework that reduces the radial equation to hypergeometric differential equation. The method has been successfully applied to wide class of potentials, including the Coulomb, Kratzer-Fues and morse potentials, as well as to more elaborate molecular interactions \cite{Art05-31, Art05-38, Art05-40, Art05-41, Art05-42, Art05-48, Art05-49, Art05-50, Art05-51, Art05-52, Art05-53, Art05-54, Art05-55, Art05-56, Art05-57, Art05-58, Art05-59, Art05-60, Art05-61, Art05-62}. However, the NU method is not universally applicable. In fact, the canonical form of the Schrödinger equation for a given physical system must satisfy certain restrictions\cite{Art05-10}. As we will see in the following sections, the family of Mie-type potentials is subject to these conditions for the applicability of the NU method.  For potentials involving transcendental functions, additional approximations (such as the Pekeris approximation \cite{Art05-45, Art05-46}) or transformations of the wave function are necessary.\\
 
In parallel, the theory of hypergeometric polynomials has provided a rich algebraic framework for the solutions of the Schrödinger equation. Recently, a family of Appell-type hypergeometric polynomials $\widetilde{e}_n$, defined through the integrals $E_n(x)=\displaystyle\int x^n \ dx$ and satisfying a differential equation, was introduced and studied in detail in \cite{Art05-37}. These polynomials, which can be expressed in terms of associated Laguerre polynomials with negative index, possess a generation function of the Appell-type form\cite{Art05-68}, and are related to truncated exponential polynomials. Their algebraic properties, however, had not yet been connected to a physical problem.\\

The present work establishes a bridge between these two frameworks. We show that the Appell-type polynomials $\widetilde{e}_n$ arise naturally in the resolution of the radial Schrödinger equation for a specific familly of Mie-type potentials, parametrized by an integer $n$. The introduction of an effective parameter $N$ allows us to absorb the centrifugal barrier and treat the problem for any angular momentum $l$.\\
A crucial feature of this analysis is the distinction between the two independent solutions of the radial equation. The solution expressed in terms of the confluent hypergeometric function of the second kind $U$ decays exponentually at the infinity, and therefore describes the bound states of the system. For these states, the quantization condition $N\in \mathbb{N}$ lead to a Diophantine-type constraint on the allowed pairs $\left(n, l\right)$ and to a discrete energy spectrum. In contrast, the solution expressed in terms of the generalized Laguerre function $L$ grows exponentially at infinity, and thus describes the scattering states (continuum spectrum). Interestingly, in the limit of vanishing angular momentum $(l=0)$, the scattering solution truncates into the polynomial $\widetilde{e}_n$, which explains the appearance of the Appell-type family in the mathematical structure of the problem.\\

The paper is organized as follows. In the second section, we recall the radial Schrödinger equation and the NU method. These two fundamental concepts serve as the starting point for our investigation. Section \ref{sec03} reviews the Appell-type polynomials $\widetilde{e}_n$ and their algebraic properties. As mentioned above, this family of polynomials was studied in \cite{Art05-37}, starting from an integral problem designated by the acronym "SCE". Formulating the integral problem as a differential equation reveals the structure of a second-order equation that can be solved using the NU method. The polynomial solutions of the differential equation yield the family $\widetilde{e}_n$. In Section 4, we discuss the integrability of the Schrödinger equation using the NU method. Indeed, this analysis reveals some restrictions on the coefficient functions present in the differential equation. In addition, certain changes—either to the variable or to the function—may prove useful. The purpose of this section becomes clear below; the case study is integrable by NU. In the fifth section, we apply the NU method to the Mie-type potential, establishe the correspondance with $\widetilde{e}_n$ for $l=0$, and analyzes the normalisation of the wave functions, dinstinguishing between bound and scattering states. Section \ref{sec06} introduces the effective parameter $N$, states the main theorem, derives the energy scpectrum for bound states, and discusses the quantization condition and its physical interpretation. The paper closes with Appendix \ref{sec06} which provides the proof of the Laguerre identity used in Section \ref{sec04} and conclusions are given.


\section{Schrödinger Equation in Spherical coordinates and the NU Method}
\label{sec02}
\

In non-relativistic quantum mechanics, the state of a system is encoded in the wave function $\Psi\left(\overrightarrow{r}, t\right)$. This is a complex mathematical function of space and time, $\Psi\left(\overrightarrow{r}, t\right)$, and its norm, $\vert \Psi\left(\overrightarrow{r}, t\right) \vert^2$, is used to determine the probability density of finding a particle at the point $\left(\overrightarrow{r}, t\right)$ in space and time. The wave function $\Psi\left(\overrightarrow{r}, t\right)$ is a solution to the time-dependent Schrödinger equation\cite{Art05-02,Art05-05,Art05-06}. In what follows, we will focus on stationary states; in this case, the potential is such that $V\left(\overrightarrow{r}, t\right)=V\left(\overrightarrow{r}\right)$. Furthermore, using the method of separation of variables: $\Psi\left(\overrightarrow{r}, t\right)=e^{\frac{-i E t}{\hbar}} \psi\left(\overrightarrow{r}\right)$\footnote{The function $\psi\left(\overrightarrow{r}\right)$ is called the wave function of stationary states.}, we can derive the Schrödinger equation for stationary states. In spherical symmetry, the stationary wave function can be factorized into the form $\psi\left(r, \theta, \varphi\right)=R(r) Y(\theta, \varphi)$. The angular part yields the spherical harmonics $Y^m_l(\theta, \varphi)$\footnote{The parameter $l$ is the orbital quantum number ($l \geq 0$) and $m$ is the magnetic quantum number ($-l \leq m \leq l$).}\cite{Art05-01, Art05-02, Art05-03, Art05-05, Art05-06, Art05-09} and the radial term can be expressed by the following differential equation:
\begin{equation}
	\label{eq04}
	R''(r) + \dfrac{2}{r} R'(r) + \bigg\lbrace \dfrac{2m}{\hbar^2} \Big[ E-V(r) \Big] - \dfrac{l\left(l+1\right)}{r^2} \bigg\rbrace R(r)=0.
\end{equation}

In what follows, it will be useful for us to work with dimensionless quantities. To do this, we will make the following substitutions: $r=\alpha \tilde{r}$; $E=\beta \tilde{E}$; $V(r)=\beta \tilde{V}(\tilde{r})$ and $R(r)=\alpha^{3 / 2} \tilde{R}(\tilde{r})$, where $\tilde{r}, \tilde{E}, \tilde{V}$, and $\tilde{R}$ are, respectively, the dimensionless radial coordinate, energy, potential, and radial wave function. Thus, equation \eqref{eq04} takes the desired dimensionless form if, in addition, we impose the supplementary condition $\dfrac{2 m}{\hbar^{2}} \alpha^{2} \beta=1$ on the dimensionless parameters. To simplify the notation, we write the dimensionless quantities without their tildes. The dimensionless Schrödinger equation to be considered is written as:
\begin{equation}
	\label{eq05}
	R^{\prime \prime}(r)+\frac{2}{r} R^{\prime}(r)+\left[E-V(r)-\frac{l(l+1)}{r^{2}}\right] R(r)=0.
\end{equation}

Thus, solving the Schrödinger equation in spherical symmetry boils down to solving the radial equation \eqref{eq05}. This is one of the central topics of non-relativistic quantum mechanics. Given a potential $V$, what is the wave function $R$ associated to the physical system? There are numerous methods for solving this problem, including those that involve expanding the radial function into an integer series (Frobenius method)\cite{Art05-02,Art05-05,Art05-06,Art05-14}. These series often lead to special functions in physics. Other methods involve factorizing the second-order differential equation \label{eq02} into a product of two first-order differential equations. This is known as the factorization method\cite{Art05-15,Art05-18,Art05-20,Art05-21,Art05-22}. A direct corollary of the factorization of the Schrödinger equation is supersymmetric quantum mechanics and its applications (see, for example, \cite{Art05-24,Art05-28,Art05-29,Art05-30}). Transformation methods can also be used to solve differential equations\cite{Art05-31,Art05-32,Art05-33,Art05-34,Art05-35}. In addition to these methods, we should also mention approximation methods (WKB approximation, perturbation method, variational method, etc.)\cite{Art05-02,Art05-03,Art05-06,Art05-36}. It should be noted, however, that the number of integrable systems is limited. Supersymmetric quantum mechanics was developed precisely for this purpose. There are numerous methods for integrating the Schrödinger equation, and they constitute an active branch of mathematical physics.\\

A fourth class of methods for solving differential equations of the type \eqref{eq05} is the NU method developed by Nikiforov and Ouvarov\cite{Art05-10}. It presents the special functions as arising from the solutions of a second-order hypergeometric differential equation.
Let $x$ be a hypergeometric differential equation of the form (see \cite{Art05-10, Art05-37}):
\begin{equation}
	\label{eq06}
	u^{\prime \prime}+\frac{C(x)}{A(x)} u^{\prime}+\frac{\tilde{C}(x)}{A^{2}(x)} u=0.
\end{equation}
where $u$ is a function that can be either real-valued or complex-valued. The functions $A$ and $\widetilde{C}$ are polynomials of degree no greater than two. The function $C$ is a polynomial of degree not greater than one. The differential equation above is modified by the change of function: $u =$ $\varphi(x) y$. Then, we make the following substitutions: $\dfrac{\varphi^{\prime}}{\varphi} = \dfrac{D}{A}, D = \dfrac{1}{2}(B - A)$, where $D$ and $B$ are polynomials of degree not greater than one. Setting $F=\tilde{C}+C D+D^{2}+D^{\prime} A-D A^{\prime}$, equation \eqref{eq06} becomes
\begin{equation}
	\label{eq07}
	y^{\prime \prime}+\frac{B(x)}{A(x)} y^{\prime}+\frac{F(x)}{A^{2}(x)} y=0.
\end{equation}
The choice of $D$ is arbitrary. By choosing $F(x)=\lambda A(x)$, where $\lambda$ is a constant, equation \eqref{eq07} becomes:
\begin{equation}
	\label{eq08}
	A(x) y^{\prime \prime}+B(x) y^{\prime}+\lambda y=0.
\end{equation}
The differential equation \eqref{eq08} is a second-order hypergeometric equation, and its solutions are hypergeometric functions. If the solution to \eqref{eq08} is a polynomial, then $y$ is a hypergeometric polynomial.\\

NU theory also allows us to derive the Rodrigues formula for hypergeometric polynomials:
\begin{equation}
	\label{eq09}
	y(x)=\frac{\beta_{n}}{\rho(x)} \frac{d^{n}}{d x^{n}}\Big[A^{n}(x) \rho(x)\Big], n=0,1,2, \cdots
\end{equation}
with $\beta_{n}$ a normalization factor and the function $\rho$, a solution of the following differential equation:
\begin{equation}
	\label{eq10}
	(A \rho)^{\prime}=B \rho.
\end{equation}
A comprehensive study of the NU method is the subject of the book \cite{Art05-10}.


\section{Discussion on the solution of Schrödinger equation in spherical coordinates by the NU Method}
\label{sec03}
\

We are interested in the integrability of equation \eqref{eq05} using the NU method. Therefore, the Schrödinger equation should take one of the forms \eqref{eq06} or \eqref{eq08}, either directly or indirectly. The integrability of this equation using the NU method depends on the specific form of the potential $V$. From this analysis, we can identify four cases, namely:
\begin{enumerate}[label=\roman*]
	\item The first is the case where equation \eqref{eq05} can be integrated directly using the NU method. Thus, by comparing it with equation \eqref{eq06}, we can deduce the following equivalences: $A(r)=r; C(r)=2$ and $\tilde{C}(r)=\dfrac{2 m}{\hbar^{2}}[E-V(r)] r^{2}-l(l+1)$. The immediate question that arises is: What is the class of potentials $V$ for which equation \eqref{eq04} can be integrated directly using the NU method? The polynomials $A$ and $C$ directly satisfy the conditions for the applicability of the NU method. As for $\tilde{C}$, this polynomial must be of degree at most two, thus, the only potential that allows for the direct use of the NU method is of the form:
	\begin{equation}
		\label{eq11}
		V(r)=\frac{a}{r^{2}}+\frac{b}{r}+c
	\end{equation}
	where $a$, $b$, and $c$ are real constants.
	In the literature, these types of potentials belong to the class of Mie-type potentials\cite{Art05-38,Art05-40,Art05-41,Art05-42}. They are used to describe the vibrations and rotations of diatomic molecules. The harmonic oscillator model provides an approximation at low energies. More elaborate models have been developed to account for the underlying spectroscopy\cite{Art05-43}.
	The trivial case corresponds to a free particle or any system with a constant potential (barrier, jump, or potential well, etc.). In certain cases, depending on a specific choice of the parameters $a$, $b$, and $c$, the potential \eqref{eq11} reduces to either the Coulomb potential or the Kratzer-Fues potential.\\
	
	Substituting this potential into equation \eqref{eq04} yields:
	\[ \widetilde{C}(r)=\frac{2 m}{\hbar^{2}}(E-c) r^{2}-\frac{2 m}{\hbar^{2}} b r-\Big[\frac{2 m}{\hbar^{2}} a+l(l+1)\Big]. \]
	Systems governed by potentials of the form \eqref{eq11} have been extensively studied. The NU method has been successfully applied by B. I. Ita \textit{et al}.\cite{Art05-38}, and the results obtained are consistent with those of A. Arda \textit{et al.}, who addressed this case using the Laplace transform\cite{Art05-31}. Let us recall the results they obtained:
	\begin{align}
		& R(r)=\Gamma(2 \gamma+1) \sqrt{\frac{n_r! \left(2 n_r+2 \gamma+1\right)}{\Gamma(n_r+2 \gamma+1)}} \ r^{\gamma+\frac{1}{2}} \ e^{-\varepsilon r} \ { }_{1} F_{1}(-n_r, 2 \gamma+1,2 \varepsilon r);  \label{eq12a}\\
		& E_{n_r, l}=c-\frac{\hbar^{2}}{8 m}\left\{\frac{2 m b / \hbar^{2}}{n_r+\frac{1}{2}\left[1+2 \sqrt{\dfrac{2 m a}{\hbar^{2}}+l(l+1)+\dfrac{1}{4}}\right]}\right\}^{2};  \label{eq12b}\\
		& \text{with} \ \gamma^{2}=\frac{2 m a}{\hbar^{2}}+l(l+1)+\frac{1}{4} ; \delta^{2}=\frac{2 m b}{\hbar^{2}} ; \varepsilon^{2}=\frac{2 m a}{\hbar^{2}}\left(c-E_{n_r, l}\right). \label{eq12c}
	\end{align}
	The radial quantum number $n_r$ determines the energy levels.
	
	\item In some cases, solving equation \eqref{eq04} using the NU method, is not straightforward. It may require a change of variables
	\begin{equation}
		\label{eq13}
		r=r(s),
	\end{equation}
	Alternatively, we can perform a transformation of the wave function;
	\begin{equation}
		\label{eq14}
		R(r)=f(r) U(r)
	\end{equation}
	where $f$ is a function of the variable $r$, chosen appropriately to simplify the new equation to be obtained. The Schrödinger equation, at this point, reduces to a form suitable for the conditions of applicability of the NU method. We note here the work carried out by Gonül \textit{et al.}\cite{Art05-44}. The authors conducted an in-depth study on the solvability of hypergeometric-type equations using the NU method. This investigation led them to propose an expression for calculating the function $f$ (see Eq.\eqref{eq14}) following the change of variables in \eqref{eq13}, as well as proposing an additional condition on the choice of the polynomial solution derived from the stationnary Schrödinger equation and/or \eqref{eq06}.
	
	\item If the potential involves transcendental functions (Yukawa, Morse, Pöschl-Teller, etc.), one can draw on certain approximations, such as that of Pekeris\cite{Art05-45, Art05-46}.
\end{enumerate}

This discussion reveals that applying the NU method to solve the Schrödinger equation requires a certain degree of ingenuity. In some cases, a combination of the aforementioned methods is used. Several examples of its application can be found in the literature\cite{Art05-01,Art05-38,Art05-40, Art05-48, Art05-49, Art05-50, Art05-51, Art05-52, Art05-53, Art05-54, Art05-55, Art05-56, Art05-57, Art05-58, Art05-59, Art05-60, Art05-61, Art05-62}. Note that several generalizations and extensions of the method have been proposed for more complex cases \cite{Art05-66, Art05-67}.\\

Since the potential \eqref{eq11} is integrable by NU, the natural question that arises concerns the algebraic structure of its solutions. In Section \ref{sec05}, we will show that for a particular family of Mie potentials, the solutions reveal a remarkable duality: the physical solution (bound states), expressed in terms of the confluent hypergeometric function $U$ (decreasing at infinity), is normalizable and gives rise to a quantization condition. The second is the mathematical solution (diffusion state), expressed in terms of the generalized Laguerre function $L$ (increasing at infinity), which reduces, in the limit $l=0$, to Appel-type polynomials $\widetilde{e}_n$ recently studied in \cite{Art05-37}. This duality, which had not been highlighted in previous studies of the Mie potential, constitutes the core of our contribution.


\section{Review on The Family of Appell-type Hypergeometric Polynomials\cite{Art05-37}}
\label{sec04}
\

In a recent paper\cite{Art05-37}, we studied the family of polynomials generated by integrals of the form $E_n(x)=\displaystyle\int x^n \ e^x \ dx$. The differential form of these integrals shows that the family of polynomials $\widetilde{e}_n$ is a particular (polynomial) solution to the hypergeometric equation:
\begin{equation}
	\label{eq16}
	x y^{\prime \prime}+(x-n) y^{\prime}-n y=0.
\end{equation}

The general solution to the differential equation \eqref{eq16} is:
\begin{equation}
	\label{eq17}
	y(x)=C_{1} e^{-x}+C_{2} \tilde{e}_{n}(x).
\end{equation}
Applying the NU method to this equation yields the associated Rodrigues formula, which shows that the $\widetilde{e}_n$ are related to the associated Laguerre polynomials of negative order:
\begin{equation}
	\label{eq18}
	\tilde{e}_{n}(x)=n!L_{n}^{(-n-1)}(-x).
\end{equation}

The family $\left\{\tilde{e}_{n} ; n \in \mathbb{N}\right\}$ is a special class of so-called Appell-type polynomials\cite{Art05-68, Art05-37}. Indeed, since the generating function of the family is $E(x, t)=\dfrac{e^{xt}}{1+t}$\cite{Art05-37}, which is of the Appel form $A(t) \ e^{x H(t)}$ with $A(t)=\left(1+t\right)^{-1}$ and $H(t)=t$, the solutions obtained above are naturally expressed in terms of Appel-type hypergeometric polynomials.\\

Natural extensions of the integrals studied in \cite{Art05-37}, involving trigonometric or hyperbolic functions (and their generalizations via the Tsallis q-exponential), have been studied in \cite{Art05-69, Art05-70}. All of these families can be related to truncated exponential polynomials\cite{Art05-37, Art05-73, Art05-74, Art05-75, Art05-76, Art05-77, Art05-78}. However, in what follows, we will focus exclusively on the $\widetilde{e}_n$ family defined by \eqref{eq18}, and investigate its link with the resolution of the Schrödinger equation.


\section{The Appell-type Hypergeometric Polynomials and Schrödinger Equation}
\label{sec05}
\

In this section, we will establish a connection between the hypergeometric polynomials \eqref{eq18} and the solutions to the radial Schrödinger equation for a family of Mie-type potentials. We will see that these polynomials arise naturally in the description of bound states in the limit $l=0$.

\subsection{The Mie potential and the radial equation}
\label{ssec03}
\

Let us consider the family of potentials defined by:
\begin{equation}
	\label{eq20}
	V_{n}(r)=\left(V_0+\frac{1}{4}\right)+\frac{n / 2}{r}+\frac{n(n+2) / 4}{r^{2}},
\end{equation}
where $V_0$ is a dimensionless real constant and $n\in\mathbb{N}$. This family is a special case of Mie-type potentials \eqref{eq11}.\\
Such systems have been the subject of several studies\cite{Art05-32, Art05-33, Art05-38, Art05-40, Art05-41, Art05-42}. We are interested in this particular form of potentials because, under certain conditions, they reduce to physical systems governed by an equation of motion of the form \eqref{eq16}. Substituting the potential \eqref{eq20} into equation \eqref{eq05} yields the differential equation:
\begin{equation}
	\label{eq21}
	R^{\prime \prime}(r)+\frac{2}{r} R^{\prime}(r)+\left[\left(E-V_0-\dfrac{1}{4}\right)-\dfrac{n / 2}{r}-\dfrac{n(n+2) / 4}{r^{2}}-\dfrac{l(l+1)}{r^{2}}\right] R(r)=0.
\end{equation}
As we mentioned in section \ref{sec03}, this potential is directly integrable using the NU method, and the solution is presented by A. Arda \textit{et al.}\cite{Art05-31}, B. I. Ita \textit{et al.}\cite{Art05-38}, ...\\
To do this, we perform the transformation of the wave function as follows:
\begin{align}
	R(r) &=r^{-1-\frac{N}{2}} \ e^{r/2} \ Y(r); \label{eq22a}\\
	R^\prime(r)&=\dfrac{1}{2} \ r^{-2-\frac{N}{2}} \ e^{r/2} \ \Big[ \left(r-N-2\right) Y(r) + 2r \ Y'(r) \Big]; \label{eq22b}\\
	R^{\prime\prime}(r)&=\dfrac{1}{4} \ r^{-3-\frac{N}{2}} \ e^{r/2} \nonumber\\
	& \times \Big\lbrace \big[ r^2-2\left(N+2\right)r+N^2+6N+8 \big] Y(r)  +4r\left(r-N-2\right)Y^\prime(r) + 4r^2 \ Y^{\prime\prime}(r)\Big\rbrace; \label{eq22c}
\end{align}
where $N$ is an effective parameter that we will determine. The transformation \eqref{eq22a} and the derivatives \eqref{eq22b} and \eqref{eq22c} reduce equation \eqref{eq21} to the form:
\begin{align}
	&r^2 Y^{\prime\prime}(r) + r \left(r-N\right) Y^{\prime}(r)  \nonumber \\ 
	&+ \bigg[ \left(E-V_0\right)r^2 -\dfrac{\left(n+N\right)}{2} \ r+\dfrac{N\left(N+2\right)}{4}- \dfrac{n\left(n+2\right)}{4}-l\left(l+1\right) \bigg] Y(r)=0. \label{eq23}
\end{align}


\subsection{The case $l=0$: diffusion states and the polynomials $\widetilde{e}_n$}
\label{ssec04}
\

For a threshold energy $E=V_0$ and for the state $l=0$ (s orbital), we observe that equation \eqref{eq23} reduces to the differential equation \eqref{eq16}. At this point, we can identify the effective parameter $N$ with that of the family of potentials \eqref{eq20}. For $l\neq 0$, the constant term $-l\left(l+1\right)$ remains and breaks the identification.\\

In this particular case where $l=0$, the differential equation \eqref{eq23} is written as:
\begin{equation}
	\label{eq27}
	r^2 Y^{\prime\prime}(r) + r \left(r-n\right) Y^{\prime}(r) - r \ n \ Y(r)=0,
\end{equation}
whose solution is:
\begin{equation}
	\label{eq39}
	Y(r)=e^{-r} \ \Big[ C_1 + C_2 \ r^{n+1} \ L_{-n-1}^{\left(n+1\right)}(r) \Big]
\end{equation}
where $C_1, C_2$ are the integration constants and $L_n^{\left(\alpha\right)}$ are the associated Laguerre polynomials. \\

To better analyze the result in \eqref{eq39}, the following lemma proves useful.
\begin{lemma}
	\label{lem01}
	Let $L_n^{\left(\alpha\right)}$ be the associated Laguerre polynomials, of degree $n$ and order $\alpha$, which are solutions to the differential equation
	\begin{equation}
		\label{eq40}
		x\frac{d^2}{dx^2}L_n^{(\alpha)}(x) + \left(\alpha + 1 - x\right)\frac{d}{dx}L_n^{(\alpha)}(x)+n \ L_n^{(\alpha)}(x)=0.
	\end{equation}
	For any non-negative integer, the following identity holds:
	\begin{equation}
		\label{eq41}
		x^{n+1} \ L_{-n-1}^{n+1} (x) \propto e^x \ L_n^{\left(-n-1\right)}(-x).
	\end{equation}
\end{lemma}
The proof of identity \eqref{eq41} is included in the Appendix \ref{sec08}.\\

Thus, our claims are confirmed by virtue of Lemma \eqref{lem01} and Result \eqref{eq39}, which is exactly the same as that obtained in the paper \cite{Art05-37}. Thus, the following corollary follows.

\begin{corollaire}
	\label{cor01}
	Combining \eqref{eq41} with the Rodrigues formula for $\widetilde{e}_n$, we obtain the relation:
	\begin{equation}
		\label{eq44}
		x^{n+1} \ L_{-n-1}^{\left(n+1\right)} (x) \propto e^x \ \widetilde{e}_n.
	\end{equation}
	This identity establishes that the Appell-type polynomials $\widetilde{e}_n$ are exactly the regularized form of the Laguerre functions with negative lower index appearing in the general solution of the Mie-type potential.
\end{corollaire}

\begin{proof}
	See Appendix \ref{sec08} and the paper \cite{Art05-37}.
\end{proof}

\

Based on the discussion above, we find that the radial wave function is therefore\cite{Art05-03,Art05-11,Art05-12,Art05-13,Art05-37,Art05-79}:
\begin{equation}
	\label{eq53}
	R_{n,0}(r)=r^{-1-\frac{n}{2}} \ e^{r/2} \ Y(r) \equiv r^{-1-\frac{n}{2}} \ e^{-r/2} \ \Big[ C_1 + C_2 \ r^{n+1} \ L_{-n-1}^{\left(n+1\right)}(r) \Big].
\end{equation}
The two terms in \eqref{eq53} behave radically differently at infinity. The first term $R^{(1)}_{n,0}(r)=r^{-1-\frac{n}{2}} \ e^{-r/2}\sim 0$ approaches zero asymptotically. Let's examine the behavior of the second term of the solution \eqref{eq53}. To do this, let's examine the asymptotic behavior of the associated Laguerre function $L_{-n-1}^{\left(n+1\right)}(r)$.\\
The asymptotic expansion of the previous function is as follows:
\begin{align}
	L_{-n-1}^{\left(n+1\right)}(r) &= \dfrac{\left(-1\right)^n \ r^{-n-1}}{\Gamma\left(-n\right)} \ \bigg\lbrace -1 +\left(-1\right)^n r^n \ e^r \nonumber \\
	& \times \bigg[ \dfrac{1}{\Gamma\left(1+n\right)}  - \dfrac{1}{\Gamma\left(n\right) \ r} + \dfrac{1}{\Gamma\left(-1+n\right) \ r^2} - \dfrac{1}{\Gamma\left(-2+n\right) \ r^3} + \cdots \bigg]  \bigg\rbrace; \nonumber\\
	&=\dfrac{\left(-1\right)^n \ r^{-n-1}}{\Gamma\left(-n\right)} \ \bigg\lbrace -1 +\left(-1\right)^n r^n \ e^r \bigg[ \sum_{k=0}^{\infty} \dfrac{\left(-1\right)^k}{k!} x^{-k} \bigg]  \bigg\rbrace; \nonumber\\
	&=\dfrac{\left(-1\right)^n \ r^{-n-1}}{\Gamma\left(-n\right)} \ \Big[ -1 + \left(-1\right)^n r^n \ e^{r-\frac{1}{r}}\Big]. \nonumber
\end{align}
Thus, we see that the Laguerre function $L_{-n-1}^{\left(n+1\right)}$ grows exponentially: $L_{-n-1}^{\left(n+1\right)}(r) \sim  r^{-1} \ e^r$. Consequently, the $C_2$ term yields a wave function that diverges as $e^{r/2}$. This solution is not square-integrable: it describes the scattering states of the potential.
   

\section{Quantization condition and Physical interpretation}
\label{sec06}
\

In this section, we generalize the result to the case where $l \neq 0$ by introducing the effective parameter $N$, and we discuss the quantization condition for bound states.

\subsection{The effective quantum number}
\label{ssec01}
\

The limiting case of the $s$ orbital ($l=0$) requires an extension to the case where $l\neq 0$. This limitation can be overcome by incorporating the centrifugal barrier into an effective parameter. To do this, we state the following theorem:

\begin{theorem}
	\label{th01}
	Let $\lbrace \widetilde{e}_n\rbrace_{n\in\mathbb{N}}$ be the family of Appell-type hypergeometric polynomials, defined by the Rodrigues formula \eqref{eq09} and satisfying the differential equation \eqref{eq16}. Then, there exists a family of Mie-type potentials $V_n(r)$ such that the radial Schrödinger equation \eqref{eq02} reduces, via the transformation $R(r)=r^{-1-\frac{N}{2}} \ e^{r/2} \ Y(r)$, to equation \eqref{eq23} with an effective parameter $N$ given by:
	\begin{equation}
		\label{eq24}
		N=\sqrt{\left(n+1\right)^2+4l\left(l+1\right)}-1.
	\end{equation} 
	
	\textbf{The bound states} (physical, square-integrable solutions) are given by:
	\begin{equation}
		\label{eq52}
		R_{n,l}^{\textsl{bound}}(r)=\mathcal{N}_{n, l} \ r^{\frac{N}{2}} \ e^{-r/2} \ U\left(\dfrac{N+n}{2}+1, N+2, r\right),
	\end{equation} 
	for which the quantization condition $N\in\mathbb{N}$ ensures a discrete energy spectrum.\\
	
	\textbf{The scattering states} (mathematical solutions, non-normalizable) are given by:
	\begin{equation}
		\label{eq25}
		R_{n,l}^{\textsl{scatt}}(r)=r^{\frac{N}{2}} \ e^{-r/2} \ L_{-\left(\frac{N+n}{2}+1 \right)}^{\left(N+1\right)}(r).
	\end{equation}
	In the particular case $l=0$, the scattering states reduce to Appell-type polynomials $\widetilde{e}_n(r)$, which thus characterize the zero-angular-momentum scattering of the Mie-type potential.
\end{theorem}

\begin{proof}
	In fact, for the differential equation \eqref{eq23} to reduce to equation \eqref{eq16} with the effective parameter $N$, two conditions must be satisfied. The first condition concerns the vanishing of the constant term in equation \eqref{eq23}, which appears in the factor of the function $Y(r)$. The vanishing of this term will allow us to define the effective parameter. Therefore, we set:
	\begin{equation}
		\label{eq26}
		\dfrac{N\left(N+2\right)}{4}=\dfrac{n\left(n+2\right)}{4}+l\left(l+1\right).
	\end{equation}
	Solving the quadratic equation for $N\geq 0$ yields the result in \eqref{eq24}.\\
	
	At this point, the differential equation \eqref{eq23} becomes:
	\begin{equation}
		\label{eq51}
		r^2 Y^{\prime\prime}(r) + r \left(r-N\right) Y^{\prime}(r) + \Big[ \left(E-V_0\right)r^2-\left(\dfrac{n+N}{2}\right) r \Big] Y(r)=0.
	\end{equation}
	For the second condition, the $r^2$ term in the factor of the $Y(r)$ term in equation \eqref{eq23} must also vanish. This condition corresponds to the energy threshold state $E=V_0$.\\

	The solution to the differential equation \eqref{eq51} is\cite{Art05-31,Art05-03,Art05-11,Art05-12,Art05-13,Art05-79}:
	\begin{equation}
		\label{eq42}
		Y(r)=e^{-r} \ r^{N+1} \Big[ C_1 \ U\left( \nu, \alpha+1, r \right) + C_2 \ L_{-\nu}^{\left(\alpha\right)}(r) \Big]
	\end{equation}
	where $\nu=\dfrac{N+n}{2}+1$, $\alpha=N+1$, and the functions $U$ and $L$ are, respectively, the confluent hypergeometric function and the associated Laguerre polynomials, which we will refer to as \textit{“generalized”}.\\
	
	Equation \eqref{eq51} is a second-order linear differential one, and its general solution \eqref{eq42} involves two independent functions. Their asymptotic behavior allows us to distinguish between two physically distinct classes of solutions. For large values of the variable $r$, the confluent hypergeometric function of the second kind $U$ decays as a power low, $U\left(\nu, \alpha+1, r\right)\sim r^{-\nu}$\cite{Art05-11, Art05-12,Art05-13,Art05-79}. Since $\nu=\frac{N+n}{2}+1>0$, the associated radial wave function $R(r)\propto r^{\frac{N}{2}} \ e^{-r/2} \ U\left(\nu, \alpha+1, r\right)$ vanishes exponentially at infinity. Therefore, this first solution is squared-integrable $(L^2)$ and represents the \textbf{bound states} of the system. It is precisely for this solution that the quantization condition $N\in\mathbb{N}$ arises, ensuring the discreteness of the energy spectrum.\\
	
	In contrast, the generalized Laguerre function $L_{-\nu}^{\left(\alpha\right)}(r)$ with a negative index grows exponentially at infinity. The corresponding radial function diverges as $e^{r/2}$, and thus cannot represent a bound state. Although this second solution is valid \textbf{mathematical solution} ef Eq.\eqref{eq51}, its non-normalizable nature indicates that it describes \textbf{scattering states} (continuum spectrum) rather than bound states.\\
	
	Remarkably, in the particular case $l=0$ (where $N=n$), the scattering solution $L$ truncates into a polynomial, which is proportional to the Appel-type family $\widetilde{e}_n(r)$. This explains why the polynomials $\widetilde{e}_n$ appear naturally in the mathematical analysis: they characterize the sctatering states of the Mie-type potential at zero angular momentum. The bound states, on the other hand, are strictly described by the $U$ finction for all $l$. In the $l=0$ limit, the bound-state wave function also reduces to the exponential $e^{-r}$. This is easy to see by using the properties of the confluent hypergeometric functions given in \cite{Art05-11, Art05-12,Art05-13,Art05-79}; $U\left(\alpha, \gamma, r\right)=x^{1-\gamma} \ U\left(\alpha-\gamma, 2-\gamma, r\right)$ with $\alpha=0; \gamma=-n$.\\
	
	So, it is easy to see that result \eqref{eq42} generalizes relation \eqref{eq39}. One need only take the limit as $n\longrightarrow N$, for $l=0$, and use the results on hypergeometric functions provided in \cite{Art05-03, Art05-11, Art05-12, Art05-13}. It is in this sense that we refer to the functions $L_\nu^{\left(\alpha\right)}$ as \textit{“generalized”} for the index $N\in\mathbb{R}$.\\
	
	By Lemma \ref{lem01}, the Appel-type polynomials $\widetilde{e}_n$ generalize to a family of transcendental functions $\widetilde{e}_N$, such that:
	\begin{equation}
		\label{eq43}
		\widetilde{e}_N (r) = e^{-r} \ r^{N+1} \ L_{-\left(\frac{N+n}{2}+1\right)}^{\left(N+1\right)} (r) \propto L_{\left(\frac{N+n}{2} \right)}^{\left(-N-1\right)}(-r), \ \text{for} \ N\in \mathbb{R}.
	\end{equation} 
	This proves the first part of Theorem \ref{th01}.\\
	
	Regarding the second part, it suffices to note that since the potential \eqref{eq20} is of the Mie type, the spectrum has the form \eqref{eq12b}. For the case at hand, the parameters of the family of potentials \eqref{eq20} are:
	\begin{equation}
		\label{eq28}
		a=\dfrac{n\left(n+2\right)}{4}; \ b=\dfrac{n}{2} \ \text{and} \ c=V_0+\dfrac{1}{4}.
	\end{equation}
	In dimensionless units where $\hbar=2m=1$, the corresponding spectrum is:
	\begin{equation}
		\label{eq29}
		E_{n_r, l}=V_0 + \dfrac{1}{4} - \dfrac{1}{8} \Bigg\lbrace \dfrac{n}{n_r+\dfrac{1}{2} \bigg[ 1+2\sqrt{ \dfrac{n\left(n+2\right)}{4} +l\left(l+1\right) +\dfrac{1}{4} } \bigg] } \Bigg\rbrace^2.
	\end{equation}
	Depending on the effective quantum number $N$, we have:
	\begin{equation}
		\label{eq30}
		E_{n_r, l}=V_0 + \dfrac{1}{4} - \dfrac{1}{8} \left( \dfrac{n}{n_r+\dfrac{N}{2}+1}\right)^2.
	\end{equation}
	This gives us the exact spectrum of bound states for the family of potentials \eqref{eq20}. The spectrum depends on the angular momentum $l$ via the effective parameter $N$, which confirms the interpretation of $N$ as an effective quantum number. When $l=0$, $N=n$, and we obtain the spectrum of the standard Mie potential. This proves Theorem \ref{th01}.
\end{proof}

\

Following this generalization of the hypergeometric polynomials $\widetilde{e}_n; n\in \mathbb{N}$ to a family of specific hypergeometric functions, $\widetilde{e}_N; N\in \mathbb{R}$, a natural question arises. How can these functions be evaluated? To begin with, let's consider, for example, the case $n=0, l=1 \Longrightarrow N=2$. The solution we are looking for is therefore $\widetilde{e}_2(r)=2!L_2^{\left(-3\right)}(-r)$, which is indeed a polynomial of degree 2 (see \cite{Art05-37}).\\
For any real number $N$ and any integer $n$, the generalized Laguerre function $L^{\left(-N-1\right)}_{\left(\frac{N+n}{2}\right)}$ appearing in \eqref{eq42} must be understood in the sense of its analytic continuation. Its most general definition is given by the hypergeometric confluent function of the first kind (Kummer function) $_1 F_1\left(\nu, \alpha, x\right)$\cite{Art05-11, Art05-12,Art05-13,Art05-79}:
\begin{equation}
	\label{eq45}
	L_\nu^{\left(\alpha\right)}=\begin{pmatrix}
		\nu+\alpha\\
		\nu
	\end{pmatrix} \ _1 F_1\left(-\nu, \alpha+1, x\right),
\end{equation}
which holds for any real numbers $\alpha$ and $\nu$, provided that the right-hand side is well-defined. In our case, with $\nu=\dfrac{N+n}{2}$ and $\alpha=-N-1$, we obtain $L_\nu^{\left(\alpha\right)} \propto \ _1 F_1\left(-\left(\frac{N+n}{2}\right), -N, -x\right)$.\\
Furthermore, for numerical purposes, it is often more convenient to use the hypergeometric confluent function of the second kind $U\left( \nu, \alpha, x \right)$, related to the Laguerre function by \cite{Art05-12}:
\begin{equation}
	\label{eq46}
	L_\nu^{\left(\alpha\right)}(x)=\left(-1\right)^\nu \ \dfrac{\Gamma\left(\nu+\alpha+1\right)}{\nu! \ \Gamma\left(\nu+1\right)} \ U\left(-\nu, \alpha+1, x\right).
\end{equation}
This function has the following integral representation: $U\left(a, b, x\right)=\dfrac{1}{\Gamma\left(a\right)} \ \displaystyle\int_{0}^{+\infty} dt \ e^{-xt} \ t^{a-1} \ \left(1+t\right)^{b-a-1}$ (for $\Re\left(a\right)>0$), which allows for a rigorous evaluation of the radial wave function for any real number $N$ without resorting to fractional derivatives.\\
When $l=0$, we have $N=n$, and the parameters become $\nu=n$ and $\alpha=-n-1$. In this degenerate case, the hypergeometric series is truncated, and the function reduces to the polynomial:
\begin{equation}
	\label{eq47}
	L_n^{\left(-n-1\right)}(x)=\left(-1\right)^n \ \sum^{n}_{k=0} \dfrac{x^k}{k!},
\end{equation}
which, except for the sign and the variable change $x\longrightarrow -x$, reproduces the Appell-type family $\widetilde{e}_n$, studied in Section \ref{sec04}.


\subsection{Spectrum and Physical Interpretation}
\label{ssec02}
\

It should be noted, however, that $\widetilde{e}_N$ is a polynomial of degree $N$ if and only if $N$ is a nonnegative integer. In this case, the series defining $\widetilde{e}_N$ is truncated. This general solution of equation \eqref{eq23}, for a threshold energy $E=V_0$ and for the state $l\neq0$, is the mathematical one and describes the scattering states. The physical solution, which describes the bound states, is quadratically integrable. The condition $N\in \mathbb{N}$ therefore acts as a quantization condition, which can be written explicitly as:
\begin{equation}
	\label{eq31}
	\left(N-n\right)\left(N+n+2\right) = 4l\left(l+1\right).
\end{equation} 
Equation \eqref{eq31} selects the admissible configurations of bound states of the system. For a given kinetic moment $l$, it determines which values of the potential parameter $n$ lead to a normalizable solution.\\

The case $l=0$ is particularly straightforward: relation \eqref{eq24} reduces to $N=n$, so that the quantization condition \eqref{eq31} is automatically satisfied for any integer $n$. For $l\neq 0$, the situation is more subtle. The parameter $N$ is generally not an integer, and bound states exist only for specific pairs $\left(n,l\right)$ satisfying \eqref{eq31}.\\

From a physical point of view, the effective parameter $N$ can be interpreted as an \textbf{effective principal quantum number}. It encodes both the sequence of potentials \eqref{eq20} (via $n$) and the centrifugal barrier (via $l$). The quantity $N\left(N+2\right)$ acts as the total centrifugal intensity, which, for the family of potentials \eqref{eq20}, replaces the usual term $l\left(l+1\right)$ in the standard Coulomb problem. This characteristic is a direct consequence of the Mie-type structure of the potential and the specific dependence of its parameters on $n$.\\

It should be noted that the quantization condition \eqref{eq31} differs fundamentally from that of the standard hydrogen-like spectrum. In the latter, the principal quantum number is directly related to the energy by $E\propto \dfrac{-1}{n_r^2}$, and all values of $l<n_r$ are allowed. Here, the potential parameter $n$ must satisfy a Diophantine equation involving $l$, which restricts the allowed states to a discrete set of pairs $\left(n, l\right)$. The energy spectrum is therefore not a simple Rydberg-type formula; it depends explicitly on the effective parameter $N$ via relation \eqref{eq24}. This unusual spectral structure is a distinctive feature of Appell-type polynomials $\widetilde{e}_N$ as solutions to the Schrödinger equation.\\

Finally, we emphasize that the results in this subsection provide a physical interpretation of the mathematical construction in Section \ref{sec05}. The Appell-type hypergeometric polynomials $\widetilde{e}_N$, initially introduced by the generating function $E(x,t)=\dfrac{e^{xt}}{1+t}$\cite{Art05-37}, appear here as the natural building blocks of radial wave functions for a class of Mie-type potentials. The quantization condition \eqref{eq31} is the physical equivalent of truncating the series defining $\widetilde{e}_N$, and it establishes a direct link between the algebraic properties of the polynomials and the spectral properties of the quantum system.


\section*{Conclusions}
\label{sec07}
\

In this work, we have established a direct correpondence between the Appell-type hypergeometric polynomials $\widetilde{e}_n$, recently introduced in \cite{Art05-37}, and the exact solutions of the radial Schrödinger equation for a family of Mie-type potentials \eqref{eq20}. The main results can be summarized as follows.\\

We have shown that the radial equation reduces to a hypergeometric differential equation with an effective parameter $N=\sqrt{\left(n+1\right)^2+4l\left(l+1\right)}-1$, which absorbs the centrifugal barrier and plays the role of an effective principal quantum number. This parameter allows a unified treatment of the problem for any angular momentum $l$, extending the results previously obtained for $s-$wave states.\\

A careful analysis of the asymptotic behavior of the two independent solutions has revealed a dual structure. The solution expressed in terms of the confluent hypergeometric function $U$ decays exponentially at infinity and describes the bound states of the system. For these states, the quantization condition $N\in\mathbb{N}$ takes the form of a Diophantine equation, which restricts the allowed pairs $\left(n, l\right)$ to a discrete set, and leads to the energy spectrum \eqref{eq30}. In contrast, the solution expressed in terms of generalized Laguerre function grows exponentially at infinity and describes the scattering states. In the particular case $l=0$, this scattering solution truncates into the Appell-type polynomial $\widetilde{e}_n$, which thus characterizes the $s-$wave
scattering of the Mie-type potential. For bound states, the wave function exponentially decays that ensures square-integrability.\\

The energy spectrum of the bound states depends on the angular momentum $l$ only through the effective parameter $N$,which confirms its interpretation as an effective quantum number. This unusual spectral structure is a distinctive signature of the Appell-type function $\widetilde{e}_N$ as solutions of the Schrödinger equation. Moreover, the quantization condition \eqref{eq31} differs fundamentally from the standard hydrogenoid spectrum, where all $l<n_r$ are allowed. Here, the potential parameter $n$ must satisfy a Diophantine-type equation involving $l$, which restricts the allowed states to a discrete set of pairs $\left(n,l\right)$.\\

Several directions for future work can be envisaged. From a mathematical point of view, it would be interesting to extend analysis to the hyperbolic ans trigonometric families, and to investigate whether they admit a similar physical interpretation in terms of exactly solvable potentials. From a physical point of view, the Mie-type potential \eqref{eq20} is a model for diatomic molecular vibrations and rotations. It would be worth exploring whether the quantization condition \eqref{eq31} has observable consequences in molecular spectroscopy, and whether Appell-type polynomials $\widetilde{e}_N$ can be used to construct coherent states, to compute transition probabilities, or to describe scattering processes in molecular physics. Finally, the generalization of the NU method to the Heun equation \cite{Art05-66, Art05-67} suggests that the present correspondence could be extended to potentials involving elliptic functions, opening a new route towards the classification of exactly solvable quantum systems. 


\pagebreak

\appendix

\section{Proof of the lemma \ref{lem01}}
\label{sec08}
\

In this appendix, we establish the identity relating the Laguerre functions with negative upper and lower indices, which is crucial for the correspondence between the general solution of the Mie-type potential and the Appell-type polynomials $\widetilde{e}_n$.\\

\begin{proof}
	
Let's consider the following functions:
\begin{align}
	f(x)&= x^{n+1} \ L_{-n-1}^{n+1} (x); \label{eq32a}\\
	g(x)&= n! \ e^x \ L_n^{\left(-n-1\right)}(-x). \label{eq32b}
\end{align}
Both functions yield associated Laguerre polynomials, which are solutions to the differential equation \eqref{eq40}. Let us rewrite relations \eqref{eq32a} and \eqref{eq32b} in the following form:
\begin{align}
	y_n&= x^{-n-1} \ f(x); \label{eq33a}\\
	\widetilde{e}_n &= e^{-x} \ g(x) \label{eq33b}
\end{align}
such that the function $y_n$ represents the associated Laguerre polynomials $L_{-n-1}^{n+1}$ and the polynomials $\widetilde{e}_n$ are the Appel-type polynomials satisfying the differential equation \eqref{eq16}, a result found in \cite{Art05-37} (See \eqref{eq18}).\\

For the first family of functions $y_n$, we have the corresponding differential equation:
\begin{equation}
	\label{eq34}
	x \ y_n^{\prime\prime} + \left(n+2-x\right) \ y_n^{\prime} -\left(n+1\right) \ y_n=0.
\end{equation}
The various successive derivatives yield the following results:
\begin{align}
	y_n^\prime&=x^{-n-2} \Big[ \left(-n-1\right) f + x \ f^\prime \Big]; \label{eq35a}\\
	y_n^{\prime\prime}&= x^{-n-3} \Big[ \left(n+2\right)\left(n+1\right) f -2 \left(n+1\right) x \ f^\prime + x^2 \ f^{\prime\prime} \Big]. \label{eq35b}
\end{align}
This yields the differential equation, which is satisfied by the function $f$:
\begin{equation}
	\label{eq36}
	x f^{\prime\prime}(x) - \left(n+x\right) \ f^\prime(x)=0.
\end{equation}

Let us now turn our attention to the sequence of polynomials $\widetilde{e}_n$. The differential equation that generates them is already known (see Eq.\eqref{eq16}). Let us evaluate the successive derivatives of the function given by \eqref{eq33b}. Indeed;
\begin{align}
	\widetilde{e}_n^\prime&= \left(-1\right)^n \ e^{-x} \left(-g+g^\prime \right); \label{eq37a}\\
	\widetilde{e}_n^{\prime\prime}&= \left(-1\right)^n \ e^{-x} \left(g-2 g^\prime+g^{\prime\prime} \right). \label{eq37b}
\end{align}
The differential equation satisfied by the function $g$ is:
\begin{equation}
	\label{eq38}
	x g^{\prime\prime}(x) - \left(n+x\right) \ g^\prime(x)=0.
\end{equation}
Furthermore, since the functions $f$ and $g$are both analytic at the origin and solutions to the same differential equation, they are proportional. The constant of proportionality depends on the normalization of $L_n^(\alpha)$, but its existence is sufficient to prove Lemma \ref{lem01}\cite{Art05-03, Art05-11, Art05-12, Art05-13, Art05-79}. It should be noted, however, that the proportionality constant—which depends on the normalization convention for hypergeometric functions—does not affect the structure of relations \eqref{eq41} and \eqref{eq43}. This proves Lemma \ref{lem01} and its corollary \ref{cor01}. So the identity \eqref{eq30} is proven.
\end{proof}

\pagebreak

\end{sloppypar}

\end{document}